# Scaling behavior of current- voltage characteristics of $Y_{1-x}Ca_xBa_2Cu_3O_{7-\delta}$ polycrystalline samples

E. Nazarova[1], K. Nenkov[2], K. Buchkov[1], A. Zahariev[1]

[1]*Institute of Solid State Physics, Bulgarian Academy of Sciences, 72 Tzarigradsko Chaussee Blvd., 1784 Sofia, Bulgaria*
[2]*IFW, Institute of Solid State and Materials Research, P.O. BOX 270017, D-01171 Dresden, Germany*

**Abstract**

I-V characteristics of polycrystalline $Y_{1-x}Ca_xBa_2Cu_3O_{7-\delta}$ samples (x=0.025 and 0.20) have been measured at different temperatures and magnetic fields in the range 0.1 T-6.9 T. The scaling behavior has been established for both samples at all magnetic fields. The dynamic exponent z displays some morphology dependence with higher value for small grain size sample $Y_{0.8}Ca_{0.2}Ba_2Cu_3O_{7-\delta}$. The static exponent ν has been determined from ρ vs. T dependence at given magnetic field. The critical exponents are field independent with one only exception (ν - for $Y_{0.975}Ca_{0.025}Ba_2Cu_3O_{7-\delta}$ sample). This is connected with the special interrelation between the vortex correlation length, ξ, and intervortex spacing α (ξ ≤ α) at all magnetic fields above $T_g$ for this sample and its better pinning.

## Introduction

Flux lines are induced by the penetrating magnetic field in superconductors in the mixed state ($H>H_{c1}$). As a result of repulsive intervortex interactions a regular Abrikosov vortex lattice is formed, similar to the crystal lattice in solids. In perfect superconductors the lattice will have a nonzero resistivity due to flux flow. However in real ceramic superconductors with high critical temperature, $T_c$, thermal fluctuations exert a strong influence on the vortex dynamics. Small coherent length, disordered pinning, anisotropy and sample dimensionality also have important contributions [1]. According to the Kim – Anderson theory [2] vortex lines move by thermal activation across the energy barriers caused by the pinning centers. A finite resistivity $\rho\sim\exp(-U/kT)$ appeared due to the vortex motion, where U is the height of pinning barrier. For low temperature superconductors U/kT is large and ρ is insignificant (except close to the $H_{c2}$ line). For high-$T_c$ superconductors U/kT is not sufficient to prevent thermally assisted flux creep and nonzero resistance (except at T=0). This concept is in consistency with experimentally observed upward curvature of current-voltage (I-V) characteristics, but can not explain the downward one at lower temperatures.

In HTS the flux line lattice is more disordered than in classical superconductors due to the small coherence length comparable to the atomic scale. Consequently a large number of pinning centers is characteristic for them [3]. In such a bulk disordered system a second order phase transition and a sharp equilibrium phase boundary between vortex-glass and vortex-liquid thermodynamic phases were proposed to exist [3-6]. The vortex-glass phase posses zero linear resistivity $\rho_{lin}=(E/J)_{J\rightarrow 0}=0$ and is regarded as "true" superconductor in contrast to the flux creep theory, where finite $\rho_{lin}$ is expected. The obtained negative curvature of I-V dependences at lower temperatures is associated with the vortex-glass phase response.

In epitaxial YBCO thin films at large fields a second order phase transition has been observed at given temperature $T_g$ [4]. According to [6] the best evidence of a true phase transition near $T_g$ is the scaling behavior existence. The I-V isotherms should collapse into two master curves above and below $T_g$, representing the liquid and glass states of the flux lines. The vortex correlation length, $\xi$ and relaxation time $\tau$ in the vortex glass state are the quantities diverging at the transition temperature $T_g$ according to the relations:

$$\xi \propto |T-T_g|^{-\nu} \text{ and } \tau \propto \xi^z, \quad (1)$$

where $\nu$ and $z$ are the static and dynamic exponents, respectively. The vortex-glass-vortex-liquid phase transition is analyzed using the scaling relation [3]:

$$E(J) \approx J\xi^{D-2-z}\varepsilon_{\pm}(J\xi^{D-1}\Phi_0/k_BT), \quad (2)$$

where J is the current density, E is the electric field, D is the dimensionality of the system under consideration and $\varepsilon_{\pm}$ are the scaling functions above and below the glass transition temperature $T_g$. After determining the proper value of $T_g(H)$ and critical exponents from the experiment the scaling should be provided. Following the described analysis experimental results of different samples have been scaled. In particular, the vortex glass transition has been observed in YBCO thin films [4, 7], single crystals [8] and polycrystalline samples [9, 10], in Bi-2212 and Bi-2223 single crystals [11, 12], thin films [13] and Bi2223/Ag tape [14].

It is important to mention that examination of the vortex dynamics in HTSC is essential not only from fundamental but from practical point of view as well. The vortex movement generates dissipation in type II superconductors, which should be controlled on the lowest possible level when the practical application of these materials is considered. The vortex-glass-vortex-liquid transition was provided in Ag-sheathed Bi-2223 tapes and critical scaling collapse was observed [14].

In this study we investigate I-V characteristics of Ca substituted polycrystalline $Y_{1-x}Ca_xBa_2Cu_3O_{7-\delta}$ samples. Recently non-power low I-V dependences have been observed in similar samples [15]. Experiments have been carried out at very small magnetic fields (up to 20 Oe) and only intergranular flux pinning is probed. We used higher magnetic fields in the range 0.1 T – 6.9 T, when the intragranular pinning is active. Chemical substitution is a successful method for pinning centers generation in HTS. The substituted atoms produced nanoregions in the material with suppressed $T_c$, and may serve as effective intragranular pinning centers [16, 17]. In particular, it has been shown that a small quantity of Ca (2-4%) increases the pinning in $Y_{1-x}Ca_xBa_2Cu_3O_{7-\delta}$ samples [17-19]. The accompanying effect of oxygen vacancy creation, especially in $CuO_2$ planes should be also important for pinning. It is known that Ca substitution leads to finer grain structure crystallization in polycrystalline $RBa_2Cu_3O_{7-\delta}$ samples [15, 17] and ruthenocuprates as well [20]. But the interrelations among the average distance between the vortex lines ($\alpha = (\Phi_0/H)^{1/2}$), the vortex correlation length, $\xi$ and grain size in the sample are important for the vortex-glass-vortex-liquid phase transition. The above discussed characteristics of Ca substituted samples make them different from the YBCO polycrystalline specimens. Establishment of scaling behavior in them will confirm the method generality and will show that overdoped Ca substituted samples in spite of their peculiarities present no exception.

## Materials and Methodology

Two polycrystalline samples with different amounts of Ca substitution ($Y_{0.975}Ca_{0.025}Ba_2Cu_3O_{7-\delta}$ and $Y_{0.8}Ca_{0.2}Ba_2Cu_3O_{7-\delta}$) have been investigated. They are prepared by the solid state reaction method from high purity $Y_2O_3$, $BaCO_3$, $CuO$ and $CaCO_3$ powders. The obtained mixture was ground. The grinding and heating steps were repeated three times. The first step was calcination at 925 °C in flowing oxygen for 23 hours. During the second step samples were heated to 930 °C for 23 hours, followed by slow cooling (2 °C/min) and additional annealing for 2 hours at 450 °C in flowing oxygen. Tablets were pressed before the third synthesis in oxygen at 950 °C for 23 hours, annealed at 450 °C for 48 hours and finally slow cooled to room temperature.

The crystal structure of the obtained specimens was examined by X-ray Powder Diffraction System STOE with $Cu_{k\alpha}$ radiation ($\lambda$=1.5405 A) at room temperature. The diffraction patterns were taken in the range $10°<2\theta<120°$ with a scan step $0.03°$.

The samples microstructure was investigated by SEM Philips 515.

The transport measurements were performed on Quantum Design PPMS. In order to prevent the sample from Joule heating effect the DC current was applied for a very short time - 0.002 sec. Thick current leads have been used, soldered to the sample's surface on big spots. The boundary values for I, V and power have been specified and the measurement was impossible in case that some of these values were exceeded. The voltages were detected with an error of several nanovolts. The applied magnetic field was perpendicular to the current direction. Sample dimensions were S= 0.116x0.211 cm$^2$, L=0.420 cm for sample with x=0.025 and S= 0.184x0.283 cm$^2$, L= 0.340 cm for sample with x=0.20, where L is the distance between the voltage leads. The electric field E, current density J and resistivity $\rho$ have been found according to the relations: E=V/L; J=I/S and $\rho$=E/J.

The values of the critical current, $J_0$, were obtained by the E-J characteristics at given temperature according to the offset criterion. The tangent was drawn to E-J curve at interception point with the electric field criterion 10 µV/cm. The critical current is determined as the current where this tangent extrapolates to zero electric field.

Two types of measurements have been performed: resistivity vs. temperature and I-V curves at constant temperatures. Both measurements were

- 3 -

done at different magnetic fields ranging from 0.1 T to 6.9 T. The values of the magnetic field chosen for any sample have been determined by previous investigations of irreversibility line performed by AC magnetic susceptibility measurements [21].

## Results and Discussion

On the Fig.1 (a, b) XRD patterns and SEM images are presented for samples with x=0.025 and x=0.20 respectively. The X-ray analysis showed that both samples have orthorhombic crystal structure without impurity phases in the limit of method accuracy. SEM investigations for $Y_{0.975}Ca_{0.025}Ba_2Cu_3O_{7-\delta}$ sample showed good crystallization with elongated grains and average grain size about 10 μm. For $Y_{0.8}Ca_{0.2}Ba_2Cu_3O_{7-\delta}$ sample grains with irregular shape and smaller average dimensions (about 5 μm) are observed.

The onset of critical temperature has been determined from temperature dependence of AC magnetic susceptibility at low field amplitude $1.10^{-5}$ T ( 0.1 Oe) and frequency f=1000 Hz. For the sample with x=0.025 $T_{conset}$=90.6 K and it is smaller when compared with no substituted YBCO. For the other sample with x=0.20 as a result of large Ca amount a significant reduction of critical temperature is observed and $T_{conset}$=81.1 K.

The transport measurements have been performed for wide temperature range at an interval of 1K at fixed magnetic fields. On the Fig.2(a, b) the ρ-J curves in double logarithmic scale are presented for both samples measured at the same magnetic field of 0.1 T. The $T_g$ temperature is also indicated. The negative curvature of ρ-J isotherms at $T<T_g$ is a signature of vortex-glass phase existence in both samples. The vortex correlation length (ξ) has been estimated

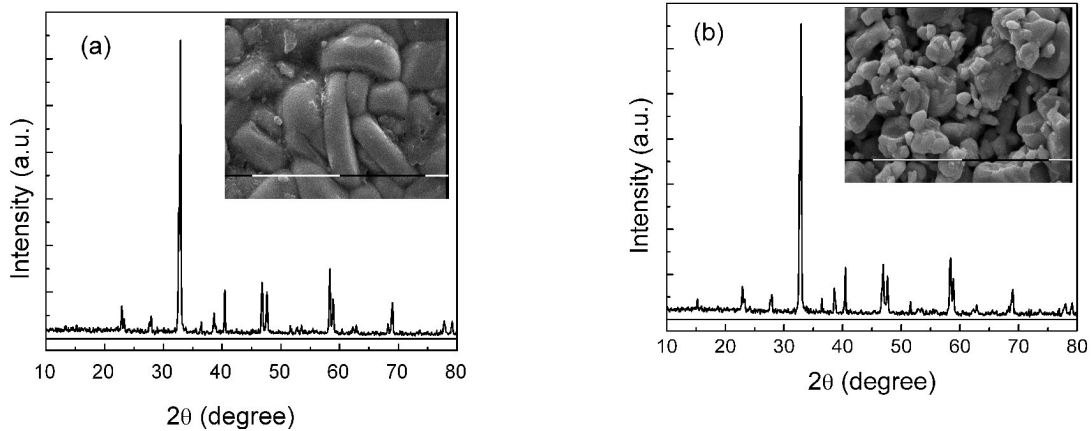

**Fig.(1).** X-ray powder diffraction patterns and scanning electron micrographs for (a) sample $Y_{0.975}Ca_{0.025}Ba_2Cu_3O_{7-\delta}$ and (b) sample $Y_{0.8}Ca_{0.2}Ba_2Cu_3O_{7-\delta}$ with magnification 5000 and marker length 10 μm.

- 4 -

by using the relation [3]

$$J_0 = k\,T/\Phi_0\xi^{(D-1)}, \quad (3)$$

where $J_0$ is the critical current at $T<T_g$, k is the Boltzmann constant and D=3. For a sample with x=0.025 at T=81 K ($T_g$=85 K) the critical current has been determined to be $4.5.10^4$ A/m$^2$ and the correlation length is about 3.5 μm. For the other sample with x=0.20 the critical current at T=62 K (also 4 K lower than $T_g$=66 K) is $6.10^4$ A/m$^2$ and the corresponding ξ value is found to be ~2.7 μm. At this field (0.1 T) inter-vortex spacing is found to be ~14.4 μm, which is about 4-5 times higher than obtained ξ values confirming the presence of vortex-glass phase below $T_g$.

Samples' dimensionality is a question under discussion [4, 22]. A 3D phase transition is accepted in a finite temperature range around the transition temperature when analytical treatments of layered system are made. Thus at $T_g$ layers are coupled at all length scales, but start to become decoupled at higher temperatures [22].

According to the model at phase transition temperature a power-low behavior is expected between voltage and current E(J, T=$T_g$) ~$J^{(z+1)/(D-1)}$ or ρ(J, T=$T_g$) ~ $J^{(z-1)/(D-1)}$ [6]. The latter dependence was used for the determination of critical exponent z. The obtained results for both samples for all magnetic fields are summarized in the Table 1.

It is seen from the Table 1 that the dynamic scaling parameter is determined within an error range of 1-2 %. z is also magnetic field independent for both samples in consistency with the vortex-glass model [3, 5].

Investigations of polycrystalline YBCO samples with different grain size show some morphology dependence of z establishing the larger z values for the samples with smaller grain size [10]. The exact values obtained for the YBCO at 0.1 T magnetic field are $T_g$=80.8 ±0.2 K and z=3.4 ± 0.33 for a sample with average grain size 6.3 μm and $T_g$=85 ±0.25 K and z=3 ± 0.4 for sample with average grain size 15.75 μm. Our results for the same field are in consistency with

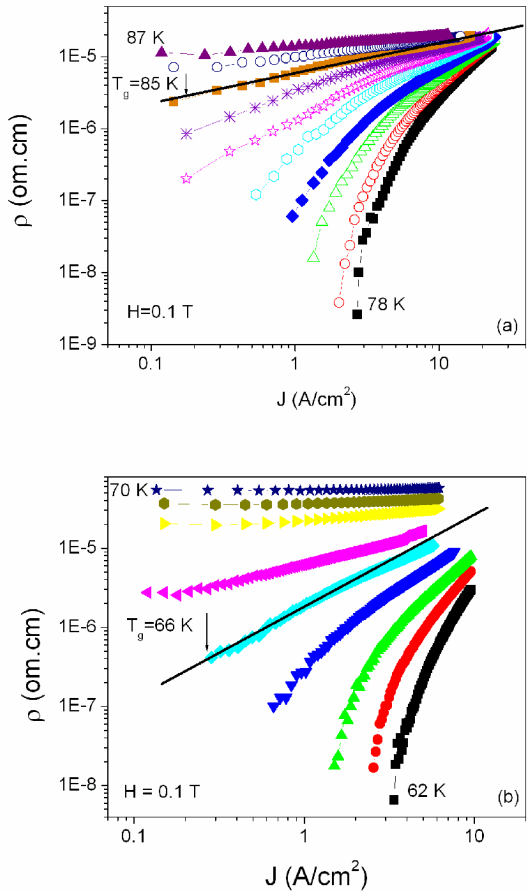

**Fig.(2).** ρ-J isotherms at 0.1 T magnetic field for (a) sample $Y_{0.975}Ca_{0.025}Ba_2Cu_3O_{7-\delta}$ at the temperature interval 78-87 K and (b) sample $Y_{0.8}Ca_{0.2}Ba_2Cu_3O_{7-\delta}$ at temperature interval 62 K-69 K at an interval of 1K. The isotherm at transition temperature is specially indicated.

Table1. Summary of critical exponent z and ν and vortex-glass-vortex-liquid transition temperature $T_g$ for both samples in selected magnetic fields, H.

| Sample | H (T) | $T_g$ (K) | z | ν |
|---|---|---|---|---|
| $Y_{0.975}Ca_{0.025}Ba_2Cu_3O_{7-\delta}$ | 0.1 | 85 | 1.84 ± 0.01 | 0.86 ± 0.03 |
| | 0.6 | 83 | 1.56 ± 0.01 | 0.81 ± 0.01 |
| | 1.69 | 79 | 1.62 ± 0.01 | 1.40 ± 0.01 |
| | 4.14 | 76 | 2.02 ± 0.02 | 1.58 ± 0.03 |
| | 6.9 | 73 | 2.04 ± 0.01 | 1.94 ± 0.02 |
| $Y_{0.8}Ca_{0.2}Ba_2Cu_3O_{7-\delta}$ | 0.1 | 66 | 3.12 ± 0.01 | 0.76 ± 0.03 |
| | 1.25 | 62 | 2.98 ± 0.02 | 0.82 ± 0.02 |
| | 2.69 | 61 | 2.48 ± 0.02 | 0.94 ± 0.01 |
| | 4.14 | 59 | 2.79 ± 0.01 | 0.95 ± 0.01 |
| | 6.9 | 56 | 3.11 ± 0.02 | 0.96 ± 0.05 |

these previously reported data when the peculiarities of the samples have been accounted.

First: comparing the z values for two investigated samples it was established that the sample with the smaller grains ($Y_{0.8}Ca_{0.2}Ba_2Cu_3O_{7-\delta}$) has the larger z (~ 3), while the sample with the larger grains ($Y_{0.975}Ca_{0.025}Ba_2Cu_3O_{7-\delta}$) shows smaller z (≤ 2). This confirms previously reported morphology dependence of dynamic exponent z [10]. Second: comparison of z values for Ca substituted and non substituted YBCO samples [10] showed that substituted samples have smaller z values for the similar grain size and magnetic fields range. The z suppression is more pronounced (about 1) for the $Y_{0.975}Ca_{0.025}Ba_2Cu_3O_{7-\delta}$ sample, while for $Y_{0.8}Ca_{0.2}Ba_2Cu_3O_{7-\delta}$ it is only several tenths. The reduction of the z exponent indicates that the vortex relaxation time is enhanced and confirms the improved pinning in Ca doped samples in comparison with non-substituted. On the other hand our previous investigation confirms that the pinning in sample with x=0.025 is better than in YBCO [21, 23]. For $Y_{0.8}Ca_{0.2}Ba_2Cu_3O_{7-\delta}$ sample the effect is smaller and combined with the significantly suppressed $T_g$ moves its irreversibility line below the YBCO..

Following the model predictions, the region of constant resistivity ($\rho_{lin}$) at small current density is investigated above $T_g(H)$. The resistivity should vanish at $T_g(H)$ according to the relation $\rho_{J\to 0} \sim (|T-T_g|/T_g)^{\nu(z-1)}$. The experimental ρ vs. T dependences at different magnetic fields, have been plotted as $\rho_{lin}$ against $(T-T_g)/T_g$ on a log-log scale. The plot is linear in the critical regime with a slope ν(z-1). Thus the ν value has been determined when the z is known. On Fig.3(a, b) the ρ vs.$(|T-T_g|/T_g)$ dependences measured at small current (20 mA) at all magnetic fields are presented for both samples. The ν values are obtained with an error of 1-5% and listed in Table 1. ν is almost field independent for $Y_{0.8}Ca_{0.2}Ba_2Cu_3O_{7-\delta}$ in consistency with the model predictions. More frequently it is close but smaller than 1. For comparison ν is 1.13-1.15 at 0.1 T for the previously discussed polycrystalline YBCO samples [10]. For sample with x=0.025 ν is in the range 0.86 -1.94 and increases more than 2 times when the field grows from 0.1T to

6.9 T. The static exponent is related to the vortex correlation length. Their possible field dependence is connected to its suppression to a value smaller than the distance between the vortices [4, 24]. We determine the $\xi$ at temperatures 2K higher than $T_g$ according to the relation (3), where J is the current density at which the resistivity starts to deviate from its constant, low-current value. For sample with x=0.20 at the highest field (6.9 T) $\xi>\alpha$. However, for the other sample $\xi\approx\alpha$ at the highest field (6.9 T) and $\xi<\alpha$ at all other fields. This difference could be the reason for a non identical behaviour of ν scaling parameter in both samples.

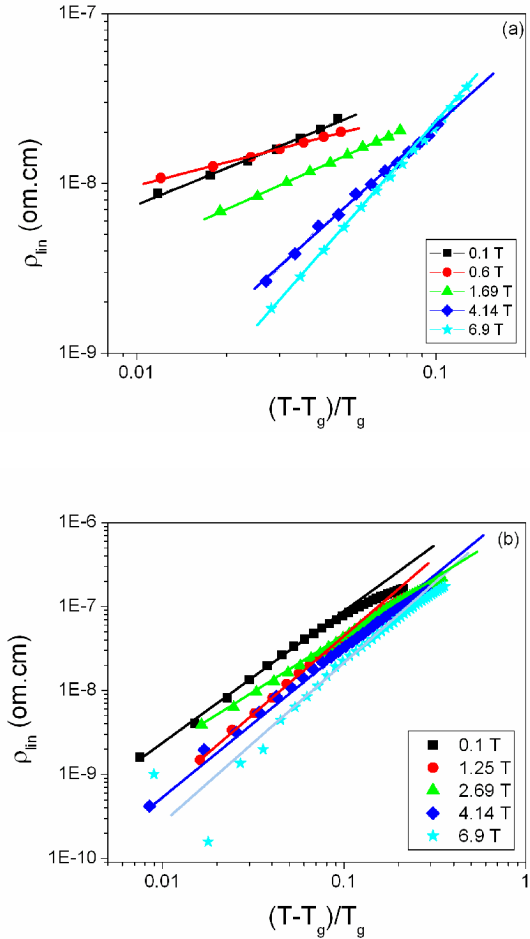

**Fig.(3)**. Low-current resistivity $\rho_{lin}$ as a function of $(T-T_g)/T_g$ for: (a) sample $Y_{0.975}Ca_{0.025}Ba_2Cu_3O_{7-\delta}$ and (b) sample $Y_{0.8}Ca_{0.2}Ba_2Cu_3O_{7-\delta}$, at different magnetic fields in the range from 0.1 T to 6.9 T.

Using the presentation on Fig.3, the width of temperature interval (ΔT) where the critical regime develops for given field is determined. It is found that this temperature interval is expanded when the magnetic field increases for both samples. However, critical regime persists at ΔT=1 K for sample $Y_{0.975}Ca_{0.025}Ba_2Cu_3O_{7-\delta}$, while ΔT =4 K for $Y_{0.8}Ca_{0.2}Ba_2Cu_3O_{7-\delta}$ at H=0.1T. At higher fields this difference became larger. For example: ΔT=7 K for sample with x=0.025 and ΔT=11.5 K for sample with x=0.20.

These measurements have been used also for another independent determination of $T_g$ value. For both samples and all magnetic fields we found good coincidence with the values determined from logρ-logJ and ρ(T) dependences. It is found that $T_g$ decreases when the magnetic field increases for both samples.

On Fig.4(a, b) the scaling behavior for both samples is presented. Fig. 4a shows the data collapse for sample $Y_{0.975}Ca_{0.025}Ba_2Cu_3O_{7-\delta}$ at 0.6T magnetic field. It presents a typical scaling behavior for both samples at low magnetic fields. Fig. 4b presents the result for sample $Y_{0.8}Ca_{0.2}Ba_2Cu_3O_{7-\delta}$ at 4.14 T. The E-J isotherms measured at temperatures T>62K scales parallel to the x axis demonstrating that J independent Ohmic regime has been reached at this conditions. Similar observations are typical for both samples at high magnetic fields where the Ohmic regime is revealed at temperatures close to $T_g$. Observation of scaling behavior is a conformation of the second order phase transition in the investigated samples.

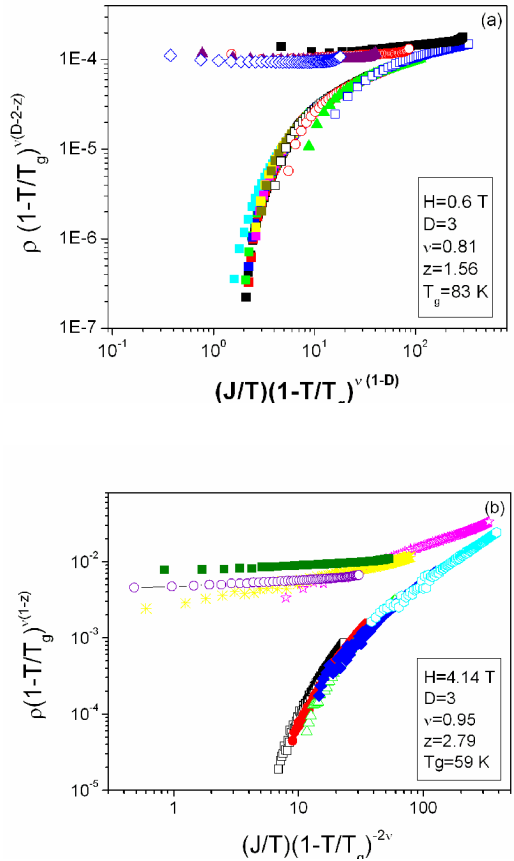

**Fig.(4)**. Scaling collapse of E-J data for sample: (a) $Y_{0.975}Ca_{0.025}Ba_2Cu_3O_{7-\delta}$ at 0.6 T in the temperature interval 70-87 K and (b) $Y_{0.8}Ca_{0.2}Ba_2Cu_3O_{7-\delta}$ at 4.14 T in the temperature interval 54-63 K.

## Conclusion

The scaling collapse of the E-J data in Ca substituted samples was established, similar to the other polycrystalline YBCO samples. Previously observed morphology dependence of dynamic exponent z was confirmed. The obtained critical parameters are field independent with one exception: ν for $Y_{0.975}Ca_{0.025}Ba_2Cu_3O_{7-\delta}$, which is connected with the special interrelation between the vortex correlation length and intervortex spacing ($\xi \leq \alpha$) for all magnetic fields above $T_g$. For Ca substituted samples the z values are smaller than usually reported for non substituted YBCO. The reasonable explanation of this fact is the better pinning, which has been established earlier independently [21, 23]. Comparison of dynamic exponents for Ca substituted samples shows better pinning in $Y_{0.975}Ca_{0.025}Ba_2Cu_3O_{7-\delta}$, resulting in enhanced relaxation time and narrow critical region.

**Acknowledgments**

The financial support of this work under the Euratom Project FU07-CT-2007-00059 is greatly acknowledged.

**References**

[1] Blatter G, Geshkenbein V. Vortex Matter In: Benneman K, Ketterson J, Eds. Superconductivity - Conventional and Unconventional Superconductors. Springer, 2008; vol.1: pp. 495-639.
[2] Anderson P, Kim Y. Hard Superconductivity: Theory of the Motion of Abrikosov Flux Lines. Rev Mod Phys 1964; 36: 39-43.
[3] Fisher M. Vortex-glass superconductivity: A possible new phase in bulk high-$T_c$ Phys Rev Lett 1989; 62 :1415-1418.
[4] Koch R, Foglietti V, Gallagher W, Koren G, Gupta A, Fisher M. Experimental evidence for vortex-glass superconductivity in Y-Ba-Cu-O. Phys Rev Lett 1989; 63: 1511-1514.
[5] Fisher D, Fisher M, Huse D. Thermal fluctuations, quenched disorder, phase transitions, and transport in type-II

superconductors. Phys Rev B 1991; 43: 130-159

[6] Huse D, Fisher M, Fisher D. "Are superconductors Really Superconducting". Nature 1992; 358: 553.

[7] Moloni K, Friesen M, Li S, Souw V, Metcalf P, McElfresh M. Universality of glass scaling in a YBa$_2$Cu$_3$O$_{7-\delta}$ thin film. Phys Rev B 1997; 56: 14784-14789.

[8] Gammel P, Schneemeyer L, Bishop D. SQUID picovoltometry of YBa$_2$Cu$_3$O$_7$ single crystals: Evidence for a finite-temperature phase transition in the high-field vortex state. Phys Rev Lett 1991; 66: 953-956.

[9] Worthington T, Olsson E, Nichols C, Shaw T, Clarke D. Observation of a vortex-glass phase in polycrystalline YBa$_2$Cu$_3$O$_{7-x}$ in a magnetic field.
Phys Rev B 1991; 43: 10538-10543.

[10] Joshi R, Hallock R, Taylor J. Critical exponents of the superconducting transition in granular YBa$_2$Cu$_3$O$_{7-\delta}$. Phys Rev B 1997; 55: 9107-9119.

[11] Safar H, Gammel P, Bishop D. SQUID picovoltometry of single crystal Bi$_2$Sr$_2$CaCu$_2$O$_{8+\delta}$: Observation of the crossover from high-temperature Arrhenius to low-temperature vortex-glass behavior. Phys Rev Lett 1992; 68: 2672-2675.

[12] Eltsev Y, Lee S, Nakao K, Tajima S. Vortex glass scaling in Pb-doped Bi-2223 single crystal. JETP Letters 2009; 90: 535-538.

[13] Yamasaki H, Endo K, Kosaka S, Umeda M, Yoshida S, Kajimura K. Current-voltage characteristics and quasi-two-dimensional vortex-glass transition in epitaxial Bi$_2$Sr$_2$Ca$_2$Cu$_3$O$_x$ films. Cryogenics 1995; 35: 263-269.

[14] Li Q, Wiesmann H, Suenaga M, Motowidlow L, Haldar P. Observation of vortex-glass-to-liquid transition in the high-T$_c$ superconductor Bi$_2$Sr$_2$Ca$_2$Cu$_3$O$_{10}$.
Phys Rev B 1994; 50: 4256-4259.

[15] Xu K, Qiu J, Shi L. Non-power-law I–V characteristics in Ca-doped polycrystalline Y$_{1-x}$Ca$_x$Ba$_2$Cu$_3$O$_{7-\delta}$. Supercond Sci Technol 2006; 19: 178-183.

[16] Koblischa M, Van Dalen A, Higuchi T, Yoo S, Murakami M. Analysis of pinning in NdBa$_2$Cu$_3$O$_{7-\delta}$ superconductors. Phys Rev B 1998; 58: 2863-2867.

[17] Huhtinen H, Awana V, Gupta A, Kishan H, Laiho R, Narlikar A. Pinning centres and enhancement of critical current density in YBCO doped with Pr, Ca and Ni. Supercond Sci Technol 2007; 20: S159-S166.

[18] Zhao Y, Cheng C, Xu M, Choi C, Yao X. Repair of grain boundaries and enhancement of critical current density by preferential chemical doping in Y-based high temperature superconductors. in A. V. Narlikar, Eds. Studies on High Temperature Superconductors, Nova Science Publisher, New York, 2002; vol.41: pp. 46-73.

[19] Laval J, Orlova T. Microstructure and superconducting properties of sintered DyBaCuO ceramics doped by Ca. Supercond Sci Technol 2002; 15: 1244-1251.

[20] Balchev N, Nenkov K, Mihova G, Pirov J, Kunev B. Superconducting properties of Ca-doped MoSr$_2$YCu$_2$O$_{8-\delta}$. Physica C 2010; 470: 178-182.

[21] Nazarova E, Zaleski A, Buchkov K. Doping dependence of irreversibility line in Y$_{1-x}$Ca$_x$Ba$_2$Cu$_3$O$_{7-\delta}$, Physica C 2010; 470: 421-427.

[22] Pierson S, Length-Scale-Dependent Layer Decoupling in Layered Systems. Phys Rev Lett 1995; 75: 4674-4677.

[23] Nazarova E, Zaleski A, Zahariev A, Buchkov K, Kovachev V, Implication of

phase separation in overdoped Y-Ca-Ba-Cu-O. J Optoelectr & Adv Matter 2009; 11: 1545-1548

[24] Roberts J, Brown B, Hermann B, Tate J. Scaling of voltage-current characteristics of thin-film Y-Ba-Cu-O at low magnetic fields. Phys Rev B 1994; 49: 6890-6894.